\documentclass[epj,nopacs]{svjour}
\usepackage{amsmath}
\usepackage{amssymb}
\usepackage{latexsym}
\usepackage[dvips]{graphicx}
\usepackage[english]{babel}

\def\bge{\begin{equation}}
\def\ene{\end{equation}}
\def\bgea{\begin{eqnarray}}
\def\enea{\end{eqnarray}}

\def\bge{\begin{equation}}
\def\ene{\end{equation}}
\def\bgea{\begin{eqnarray}}
\def\enea{\end{eqnarray}}

\def\ls{\raise 1.5pt\hbox{$\,<\;$}\kern -10.5pt\lower3.5pt
          \hbox{$\sim$}\kern 1.5pt} 
\def\gs{\raise 1.5pt\hbox{$\,>\,$}\kern -9.5pt\lower3.5pt
          \hbox{$\sim$}\kern 1.5pt} 

\newcommand{\mi}{m_{\rm i}}
\newcommand{\mbb}{m_{\rm bb}}
\newcommand{\mg}{m_{\rm g}}
\newcommand{\me}{m_{\rm e}}

\usepackage{color}

\usepackage{ulem} 

\begin{document}
\sloppy
\title{The Origin of Rest-mass Energy}
\author{Fulvio Melia\thanks{John Woodruff Simpson Fellow.}}
\institute{Department of Physics, the Applied Math Program, and Department of Astronomy, \\
              The University of Arizona, Tucson, AZ 85721,
              \email{fmelia@email.arizona.edu}}

\authorrunning{Melia}
\titlerunning{The Origin of Rest-mass Energy}

\date{\today}

\abstract{Today we have a solid, if incomplete, physical picture of how inertia is created
in the standard model. We know that most of the visible baryonic `mass' in the Universe
is due to gluonic back-reaction on accelerated quarks, the latter of which
attribute their own inertia to a coupling with the Higgs field---a process that
elegantly and self-consistently also assigns inertia to several other particles.
But we have never had a physically viable explanation for the origin of rest-mass
energy, in spite of many attempts at understanding it towards the end of the
nineteenth century, culminating with Einstein's own landmark contribution in his
Annus Mirabilis. Here, we introduce to this discussion some of the insights we
have garnered from the latest cosmological observations and theoretical modeling
to calculate our gravitational binding energy with that portion of the Universe to
which we are causally connected, and demonstrate that this energy is indeed equal
to $mc^2$ when the inertia $m$ is viewed as a surrogate for gravitational mass.}
\maketitle

\section{A Brief History of $E=mc^2$}
Today we take it for granted that a particle with {\it inertia}, $m_i$, carries an irreducible 
amount of energy---even when at rest with respect to the observer---given by Einstein's famous 
formula, $E=\mi c^2$. Every object gains kinetic energy, $K$, under the accelerating influence 
of an external force, and it loses potential energy, $\Phi$, when allowed to {\it fall} 
freely in a region where it experiences an attraction to something else. No matter how 
$K$ and $\Phi$ change, however, the rest-mass energy $E=\mi c^2$ is an immutable feature of 
$\mi$. So why should inertia, which has no obvious connection to $K$ and 
$\Phi$, be associated with energy, and why is it possible for $E$ to be converted back
and forth into $K$ and/or $\Phi$ when $\mi$ is allowed to change, e.g., via the annihilation
of a particle-antiparticle pair? 

Contrary to conventional wisdom, Einstein was not the first to consider the possible 
conversion of `mass energy' into other forms of energy, and actually did not formally prove 
their equivalence either. In 1881, the future Nobel laureate J. J. Thomson realized that---when
viewed as a charged sphere---an electron moving through an `aether' resists being accelerated
more than a similarly uncharged object \cite{Thomson:1881}. Much earlier, G. G. Stokes had
drawn similar conclusions in the context of hydrodynamics, showing in 1844 that a body's
inertia increases when moving through an incompressible perfect fluid \cite{Stokes:1844}.
Quite remarkably, both of these explanations for the origin of inertia would eventually
constitute a historical echo of the Higgs mechanism (see \S~III below), proposed more than
a century later, though based on a surprisingly similar idea \cite{Englert:1964,Higgs:1964}.

Thomson viewed this effect as arising from the electromagnetic field carried by the charge
itself, so he assigned to it an effective momentum and an apparent {\it electromagnetic mass}.
At least part of the mass of the electron could thus be viewed as arising from its electromagnetic 
self-energy---requiring some sort of equivalence between inertia and energy. Over the next two
decades, this idea was fleshed out in considerable detail by O. Heaviside \cite{Heaviside:1889},
G.F.C. Searle \cite{Searle:1897}, M. Abraham \cite{Abraham:1903} and H. A. 
Lorentz \cite{Lorentz:1892,Lorentz:1904}. Its development proceeded to the point where the 
radiation reaction force, ${\bf F}_{\rm em}$, acting on a charged particle due to the momentum 
and energy carried away by the radiation it produces, could be formally incorporated into the 
Abraham-Lorentz equation \cite{Melia:2001}.

It had been known since 1884, when J. H. Poynting published \cite{Poynting:1884} his now famous 
theorem on the conservation of energy in an electromagnetic field, that Maxwell's equations contained
the ingredients necessary to calculate both the energy and momentum density carried by a radiation
field.  The relationship between these two dynamical attributes, together with the Larmor equation 
yielding the rate of energy loss by an accelerated charge, could therefore be used to infer the 
particle's momentum loss rate, from which one could see that \cite{Melia:2001}
\begin{equation}
{\bf F}_{\rm em}\equiv {2q^2\over 3c^3}\ddot{\bf v}\;,\label{eq:radreaction}
\end{equation}
where $q$ is the particle's electric charge. One could thus interpret from this that
the field has an effective mass 
\begin{equation}
m_{\rm em}\equiv {2q^2\over 3c^3}{1\over\tau}\;,\label{eq:mem}
\end{equation}
yielding ${\bf F}_{\rm em}\equiv m_{\rm em}\dot{\bf v}$, if one identifies $\tau\equiv r_q/c$
as the light travel-time across the radius $r_q$ of the charge---a reasonable estimate of
the time associated with dynamical changes in $\dot{\bf v}$. And given that the electric
self-energy of a charge $q$ spread evenly across the surface of a sphere of radius $r_q$ is
\begin{equation}
E_{\rm em}={1\over 2}{q^2\over r_q}\;,
\end{equation}
one immediately infers the implied equivalence of the field's energy and effective mass:
\begin{equation}
E_{\rm em}\equiv {3\over 4}m_{\rm em}c^2\;.\label{eq:E43}
\end{equation}
Of course, this electromagnetic mass requires a particle to be charged, so it could not
apply to everything. Nevertheless, one cannot but marvel at the strong similarity between 
Equation~(\ref{eq:E43}) and Einstein's formula $E=\mi c^2$. And this first formal attempt 
to find an equivalence between mass and energy preceded special relativity by several decades.

Following other developments in finding the `correct' relationship between mass and 
energy, Fritz Hasen\"ohrl created in 1904 a thought experiment involving the heat (i.e., 
`blackbody') energy inside a moving cavity \cite{Hasenohrl:1904,Hasenohrl:1905}. As we shall 
see shortly, Einstein's own derivation of the relationship between inertia and energy was 
based on very similar physics. Hasen\"ohrl published several different versions of his
argument, but one can appreciate the gist of his thought experiment by simply considering 
the first \cite{Hasenohrl:1904}. He imagined filling a perfectly reflecting cavity
with `heat,' i.e., blackbody radiation, emitted symmetrically at the two ends of a cylindrical
container. Since identical radiation (or photons, in modern parlance) is emitted at each
end according to an observer sitting inside, the external forces applied to counter the
radiative reaction forces (analogous to Eq.~\ref{eq:radreaction}) are equal and opposite.

But to an observer sitting in the laboratory, watching the same cavity moving past them
at constant velocity, {\bf v}, the radiation emitted in the direction of {\bf v} is 
Doppler blue-shifted, while that emitted in the opposite direction is red-shifted.  
And since blue-shifted photons carry more momentum than their red-shifted counterparts, 
the two external forces seen in the laboratory must now be different in order to maintain
the cavity moving at constant velocity. Hasen\"ohrl applied the classical work-energy
theorem, equating the net difference in work exerted by the external forces to the
change in the cavity's kinetic energy, to show that the blackbody radiation has an 
equivalent mass $\mbb=(4/3)E_{\rm bb}/c^2$. Actually, his first publication erroneously quoted 
this result as $\mbb=(8/3)E_{\rm bb}/c^2$, but he corrected his algebraic mistake in a subsequent
paper after receiving communication from M. Abraham.

The importance of this step was Hasen\"ohrl's extension of the result in Equation~(\ref{eq:E43})
to non-charged particles. Indeed, as we shall see shortly, his thought experiment was
very similar to that of Einstein, which was published the following year. One may thus wonder
why his expression contained the factor $4/3$ instead of simply $1$. As it turns out,
this was not due to his use of classical physics, as one might suspect but, rather,
to the fact that he incorrectly ignored the mass being lost by the cylinder's caps 
while they are emitting heat \cite{Boughn:2011}.

Such was the impact of Hasen\"ohrl's argument, however, that even as late as 1909, Max 
Planck \cite{Planck:1909} included in one of his lectures the statement ``that the 
blackbody radiation possesses inertia was first pointed out by F. Hasen\"ohrl." But 
the correct answer, of course, was published by A. Einstein \cite{Einstein:1905} in 
one of his four Annus Mirabilis papers of 1905. Couched in the language of special 
relativity, Einstein's argument was---in retrospect---remarkably simple though, in 
the end, he approximated away the relativistic parts anyway, so his answer is 
derivable purely from classical physics, based on the Doppler effect. 

Einstein considered a single point particle, moving with velocity {\bf v} in the
laboratory frame, radiating away a quantity of energy $\Delta E^\prime$ with 
front-back symmetry in its own rest frame. For simplicity, he assumed that $\Delta E^\prime/2$ 
is radiated in a direction parallel to and anti-parallel to {\bf v}. According to the
relativistic Doppler-shift formula, an observer in the laboratory sees the radiation
carrying away an energy $(\Delta E^\prime/2)\gamma(1+\beta\cos\theta^\prime)$, where 
$\theta^\prime$ is the angle between {\bf v} and the direction of propagation of the radiation,
and $\gamma\equiv 1/{\sqrt{1-\beta^2}}$ is the Lorentz factor in terms of $\beta\equiv 
|{\bf v}|/c$. Thus, the difference in kinetic energy of the particle between the laboratory 
and rest frames is simply 
\begin{equation}
\Delta K-\Delta K^\prime=\Delta E^\prime(\gamma-1)\;.
\end{equation}
In the low-velocity limit, where the relativistic parts are approximated away, this 
equation becomes
\begin{equation}
\Delta K-\Delta K^\prime={1\over 2}{\Delta E^\prime\over c^2}v^2\;.
\end{equation}
He then argued that since the particle is giving away an amount of energy $\Delta E^\prime$,
its mass must have diminished by $\Delta \mi=\Delta E^\prime /c^2$. 

It is important to note, however, that Einstein made several sweeping conclusions from
this result, including (i) that it applies to all bodies and all forms of energy,
and (ii) that it remains true even at higher velocities (where relativity would indeed
introduce corrections to the classical outcome). But he never actually proved
any of these claims, even in the subsequent handful of papers he published on this topic
over the next several decades. Today we know this result is correct because it has been
verified experimentally to incredible accuracy. It has never been proven theoretically,
however, and {\it the fundamental reason why inertia ought to be associated with energy 
has remained a complete mystery to this day}.

\section{Inertia and Gravitational Mass}
To properly address the question of why a `rest mass' $\mi$ represents an energy $\mi c^2$, 
we first need to refine and clarify our concepts of inertia and gravitational charge, 
which we shall call $\mg$ to properly distinguish it from $\mi$. Newton viewed inertia 
to be a conserved and irreducible property of matter, and did not consider $\mi$ and $\mg$ 
to be distinct \cite{Newton:1687}. By `inertia' we shall strictly refer to the 
proportionality constant between an applied force and an object's consequent acceleration, 
according to Newton's second law of motion,
\begin{equation}
{\bf F}=\mi{\bf a}\;.\label{eq:Newton2}
\end{equation}
The quantity $\mi$ retains this meaning in relativity, where it is considered to be the
inertial mass in the object's {\it rest} frame. Since an observer in this frame can reduce 
their equation of motion to the classical limit shown in Equation~(\ref{eq:Newton2}), they 
could with equal validity refer to $\mi$ as either the object's inertia or its `rest 
mass' $m$. No doubt, this is a very basic concept, but we need to be clear that `inertial mass' 
strictly represents an object's resistance to acceleration when a force is applied to it 
in the classical limit.

Gravitational charge, on the other hand, arises in the context of Newton's universal
law of gravitation,
\begin{equation}
{\bf F}_1=-{Gm_{{\rm g}1}m_{{\rm g}2}\over r^2}\hat{\bf r}\;,\label{eq:grav}
\end{equation}
expressing the force ${\bf F}_1$ experienced by particle 1 (with gravitational charge
$m_{{\rm g}1}$) due to the gravitational influence exerted by particle 2 (with gravitational
charge $m_{{\rm g}2}$). The radius vector ${\bf r}=r\hat{\bf r}$ points from 2 to 1, 
and we have explicitly included a negative sign in this equation, arising from the fact that 
gravity is {\it always} attractive---a feature that will shortly become highly relevant 
to our discussion concerning the relationship between $\mi$ and $\mg$. The quantity $G$ 
is the `gravitational' constant, whose numerical value and physical units depend on 
how we {\it choose} to define $\mg$, say in terms of the (dimensionless) number of 
atoms in an object, or its inertial mass $\mi$ in kilograms. The conventional value
of $G$ that we are all familiar with arises when we force the equality $\mi=\mg$.

The latter possibility---that $\mi$ and $\mg$ might be related, perhaps even equal---arises 
from the observation that they both represent the amount of `something' in the object. 
Certainly, at the time of Newton, there weren't too many options to consider. If one were 
to double the {\it quantity of matter}, as Newton would have put it, one would reasonably 
expect from simple experimentation that its inertia would also double. Likewise, 
doubling the quantity of matter in object 1 would double the gravitational force in 
Equation~(\ref{eq:grav}). Today we know much more and realize at a very fundamental 
level that these two `quantities' need not be the same physically.  For example, if we 
were to naively stick two identical objects together, we could double the attribute that 
gives rise to inertia, while also doubling the analogous (but different) attribute 
responsible for the gravitational charge. In both cases, $\mi\rightarrow 2\mi$ and 
$\mg\rightarrow 2\mg$, even though $\mi$ and $\mg$ might have nothing to do with 
each other. In the absence of any more definitive information, the best one could
argue is therefore that $\mi\propto \mg$, certainly not that $\mi=\mg$. But even this
statement is fraught with peril given what we now know about the `equivalence' of mass 
and energy and the fact that, in general relativity, the spacetime curvature really
responds to energy, not mass, as we shall discuss later in this paper. Nonlinear effects 
that increase the self- (or binding) energy of an object as its gravitational charge 
increases may therefore destroy the simple constancy of $\mi/\mg$ if inertia is unrelated 
to gravity \cite{Yarman:2019}.

But at least in this regard, experimentation does provide us with a very firm indication
that $\mi$ remains proportional to $\mg$ over all the scales that have been tested thus far. 
Most of the experiments attempt to compare the acceleration 
of two laboratory-sized objects of different composition in the presence of an external 
gravitational field. Many high-precision E\"otv\"os-type of measurements have been
made, starting with the pendulum experiments of Newton and Bessel, to the classic
torsion-balance version of E\"otv\"os \cite{Eotvos:1922}, Dicke \cite{Dicke:1969} and
others. In the latest version of these torsion-balance experiments, two objects of 
different composition are rested on a tray and suspended horizontally by a fine wire.
For example, the `E\"ot-Wash' experiments have used such devices at the University
of Washington to compare the accelerations of various materials toward movable 
laboratory masses, the Sun and the galaxy \cite{Su:1994,Baessler:1999}, reaching a
relative precision \cite{Wagner:2012} of $2\times 10^{-13}$. (For a recent review, 
see Tino et al. \cite{Tino:2020}.) Another way to say this is that, as far as we can
tell, everything in an object that gives rise to inertia also contributes 
{\it proportionately} to its gravitational charge.

As is well known, this proportionality between $\mi$ and $\mg$ is the basis for Einstein's 
principle of equivalence. One can easily understand this from Equations~(\ref{eq:Newton2})
and (\ref{eq:grav}), which show that particles ($j=1...n$)---much closer to each other than
the scale over which a gravitational field is changing---are all accelerated at an equal rate 
proportional to the constant $m_{{\rm g}j}/m_{{\rm i}j}$. An observer could therefore not distinguish 
this situation from an analogous one in which they were being observed in a local, non-inertial 
frame accelerating uniformly in the opposite direction. 

So why couldn't this equivalence apply to other forces as well? For example, why couldn't
we argue that the amount of charge in an object is proportional to its matter content? Then
the Coulomb force acting on it analogously to Equation~(\ref{eq:grav}) would be proportional
to its net charge, $q_1$. Doubling the quantity of matter would result in $q\rightarrow 2q$
and $\mi\rightarrow 2\mi$, so that the ratio $q/\mi$ would always remain the same. In this
case, we would see an equivalence between inertia and the electric charge, perhaps leading
us to propose an alternative equivalence principle based on the notion that we could not
distinguish between charges accelerated in an electromagnetic field and the analogous
situation of charges being viewed in a non-inertial frame uniformly accelerated in the
opposite direction.

The answer, of course, is that the other forces all lack the unique combination of properties
that allow gravity to function in this way. Gravity has a single charge, unlike 
electromagnetism which has two, or quantum chromodynamics which has three (red, green and blue) 
and the corresponding antiquark colors. So gravity is always attractive, while the others 
can vary depending on the charge balance. In addition, gravitational charge cannot be
annihilated, so that all forms of energy have an effective $\mg$ that accumulates, as does
inertia, while electric charge can be completely removed from an object. In other words, 
gravity is the only force for which the proportionality between its charge and $\mi$ is 
guaranteed. And equally important, it is the only force for which one may reasonably expect 
its charge to extend its influence over a vast volume of space (i.e., the cosmos). In spite 
of the fact that the Coulomb force is itself an inverse-square law, it is energetically
prohibitive to maintain a separation of charges over distances extending beyond the
laboratory or, in the most extreme situation, beyond the magnetosphere of a pulsar,
smaller than a typical city here on Earth. The Universe is therefore neutral 
on large scales---specifically because the electromagnetic force contains more than 
one charge. So the equivalence principle could only work for gravity, and we are led
to the conclusion that inertial mass must therefore be proportional to the gravitational 
charge, which we shall henceforth sometimes call the `gravitational mass.' And to 
simplify the discussion even further, we shall often `choose' the relevant constants
(such as $G$) to have values and units that allow us to set the inertial mass and 
gravitational charge equal to each other, thereby defining the {\it rest mass}, $m\equiv\mi=\mg$.

\section{The Higgs and QCD Inertia}
Without unduly preempting our discussion in \S~V, the obvious question arising from
the conclusion in the previous section centers on the issue of whether rest-mass energy, 
$mc^2$, can really be associated with the object's inertia, $\mi$, or whether it is in 
fact an energy due to a physical influence involving its gravitational charge, $\mg$. We 
would not be able to tell the difference since $\mi/\mg=$ constant, which permits inertia 
to act as a {\it surrogate} for $\mg$. In that case, it wouldn't even matter what the 
origin of inertia were, as long as we could identify the physics that generates an energy 
$\mg c^2\rightarrow mc^2$ (which we shall do in \S~V). Nevertheless, for the sake of clarity 
and completeness, we shall here first summarize the current situation concerning 
the origin of $\mi$.

In Newton's view of the world, inertia was an intrinsic property of matter,
manifested by objects moving relative to an absolute space. But several early thinkers 
following Newton, notably Berkeley \cite{Berkeley:1710} and Mach \cite{Mach:1872}, already 
questioned an independently defined absolute space, and instead proposed that inertial frames 
are those that are unaccelerated relative to the `fixed stars' or, more accurately, relative 
to a carefully defined mean of all the matter in the Universe.  Einstein called this `Mach's 
principle' and considered it to be foundational in the development of his general relativity 
theory, but he eventually realized that these two are actually incompatible with each other 
\cite{Einstein:1917,Einstein:1946}. Though the equivalence principle had suggested to him that 
inertia must be due to the gravitational influence of the whole Universe, Einstein eventually 
realized that this influence disappears completely for a particle in free-fall. While the 
particle experiences zero gravity in this frame, it nevertheless still exhibits inertial 
properties.

Mach himself never explicitly stated how or why his view of inertia ought to be formalized 
as some kind of new physical law, so he never provided a physical mechanism describing how
the distant matter in the Universe affects the motion of a local particle. But Mach's
principle has been invoked many times in the development of alternative gravity theories.
For example, Dennis Sciama attempted in 1953 \cite{Sciama:1953} to express Mach's principle 
in more quantitative terms by proposing the addition of an acceleration-dependent contribution
to Newton's law of gravity (Eq.~\ref{eq:grav}). Sciama called this effect an `inertial 
induction.' Later, Brans and Dicke \cite{BransDicke:1961} incorporated Mach's principle into
an alternative theory to general relativity, by setting up a framework in which the
gravitational constant $G$ is determined by the structure of the Universe. In their
approach, the unit of inertial mass is taken to be the Planck mass (i.e., $m_{\rm P}^2\equiv
\hbar c/G$), so that a changing mass results from a changing $G$, which in
turn can be viewed as the Machian consequence of a changing Universe.

But in spite of these attempts at physically interpreting inertia as an effect due
to distant matter in the Universe, the situation today regarding Mach's principle is
perhaps best summarized by Abraham Pais \cite{Pais:2005}: ``It must be said that, as 
far as I can see, to this day, Mach's principle has not brought physics decisively farther. 
It must also be said that the origin of inertia is and remains the most obscure subject in 
the theory of particles and fields." Quite remarkably, though, at least a partial answer
appears to have been found in the intervening period.

In ordinary matter, ignoring for brevity and simplicity other possible issues associated
with dark matter and dark energy in a cosmological context, inertia is overwhelmingly dominated
by the nuclei, $\mi\sim m_{\rm N}$, specifically, protons and neutrons. Electrons are far smaller 
($\me<m_{\rm N}/1000$) and---if we take the liberty of borrowing the $E=mc^2$ result to convert 
the nuclear binding energy into an effective inertial mass---other contributions to the mass 
of the nucleus are but a small fraction of $m_{\rm N}$ (typically less than 1 percent). Thus,
to understand the origin of atomic inertia and, by extension, most of the inertia of
ordinary matter in the Universe, one must uncover the origin of proton and neutron 
masses and, to a lesser extent, the origin of electron mass.

Today, the standard model of particle physics is well established and experimentally
confirmed. It encompasses electromagnetism, the weak force and strong interactions, and 
provides a self-consistent classification of all the known elementary particles. It is 
nevertheless still incomplete because it does not (i) include gravity, (ii) account 
for baryon asymmetry and dark matter and (iii) allow for the inclusion of dark energy, 
if the latter turns out to be something other than a cosmological constant, $\Lambda$. 
Some of the key steps in its development have been (i) the unification of the electromagnetic 
and weak interactions by Glashow \cite{Glashow:1961}, (ii) the incorporation by Weinberg and 
Salam of the Higgs mechanism to generate inertial masses for some of its 
particles \cite{Englert:1964,Higgs:1964,Weinberg:1967,Salam:1968} (more on this below), 
and (iii) the discovery of various new particles it predicted, such as the $W^\pm$, 
$Z^0$ and Higgs bosons (see, e.g., Oerter \cite{Oerter:2006} for a detailed review).

Its structure contains six quarks (fermions that carry color charge), which are used in
various combinations to form the meson and baryon hierarchy; six leptons (including 
electrons and neutrinos); twelve spin-1 gauge bosons that mediate the strong, weak, 
and electromagnetic interactions; and one spin-0 scalar boson, i.e., the recently 
discovered Higgs particle. The gauge bosons include the aforementioned $W^\pm$ and $Z^0$ 
carriers of the weak force, as well as the massless photon responsible for the 
electromagnetic interaction. The remaining eight gauge bosons are various color 
combinations of gluons that mediate the strong force inside mesons and baryons, 
such as the proton and neutron.

The quark, electron, $W^\pm$ and $Z^0$ inertial masses are generated via the Higgs
mechanism that we shall discuss shortly. The proton and neutron masses, however, are
much larger than the mere sum of their enclosed quark and gluon fields. As surprising
as it may seem, it is actually possible to measure individual quark masses based on
the reconstruction of jets they induce in high-energy collisions. This is the method
used to measure the top quark mass, while the bottom and charm masses may also be
inferred from the mass of meson resonances, such as bottomonium and charmonium,
since these appear to be non-relativistic quark-antiquark bound states. The other
three light quark masses (strange, down, up) may be inferred from the spectroscopy of
low-lying pseudoscalar mesons, such as $\pi$, $K$, and $\eta$, whose inertial masses
depend sensitively on the light-quark masses.

As noted earlier, however, this beautiful, self-consistent picture does not yet explain why 
the nucleon mass is $\sim 20$ times larger than the sum of the quark masses within it. 
Ironically, this is where the highly original development concerning the electromagnetic 
mass in the nineteenth century resurfaces (see \S~I above), notably via arguments of the form 
expressed in Equations~(\ref{eq:radreaction}) and (\ref{eq:mem}). That proposal was based on 
the idea that energy and momentum carried away by the electromagnetic field provided a back-reaction
on the radiating particle being accelerated, thereby generating inertia. There are several
fundamental reasons from quantum electrodynamics why this mechanism cannot work for the electron,
in part because this mechanism produces infinite multiplicative factors representing the mass. 
Remarkably, however, a very similar approach does work in quantum chromodynamics. Detailed 
calculations from first principles have shown that most ($\sim 95\%$) of the nucleon's inertia 
is generated by the back-reaction of color gluon fields resisting the acceleration of quarks
and (the similarly colored) gluons inside the baryons \cite{Durr:2008}. Actually, this process
accounts very well for most of the inertia in the entire low-lying meson and baryon 
distribution.

Most of the inertial mass in ordinary matter can therefore be understood as arising from
the back reaction of gluons on the quarks that radiate them in response to the acceleration
they are subjected to by external forces. This is a rather profound statement because it tells
us that inertia originates dynamically, principally to conserve momentum, rather than from 
some Newtonian definition of irreducible internal `mass.' It should now become clearer why
the statement made at the beginning of this section is so essential to this whole
discussion. Attempting to assign `rest energy' to inertia---when viewed as an emergent
property---doesn't make much physical sense.  Instead, interpreting inertia as a surrogate 
for how much `$\mg$' a quark (say) possesses allows us to pursue a more physically meaningful 
investigation of how gravitational charge is involved in the generation of energy.

The story is not yet complete, however, because individual quarks and some of the leptons
and gauge bosons also have inertial mass, which must be due to something else. 
Conventional wisdom today has it that this type of inertia, distinct from the one generated
by the QCD interactions discussed above, is due to a coupling of these particles to a 
pervasive spin-0 scalar field \cite{Englert:1964,Higgs:1964} known as `Higgs.' Much has been 
written about this mechanism \cite{Cheng:1984}, and the discovery of the Higgs boson itself 
appears to have cemented our basic understanding of how inertia is generated for particles 
in the standard model that would otherwise have to remain massless in order to satisfy several
required symmetries. The way this mechanism works is rather easy to explain, but it also
contains an important caveat that will leave us wondering whether we have actually uncovered
the whole truth.

All of the particles in the standard model (in the absence of a Higgs field, $\Phi$) must 
have zero mass in order to comply with various (presumed) symmetries. The Lagrangian density 
representing gauge bosons, for example, cannot contain `mass' terms, such as $m_w W^\pm_\mu 
W^{\pm\,\mu}$, which would violate gauge invariance. In physics, we measure distances and times, 
velocities and acceleration in order to infer the particle dynamics. But the latter results from 
`forces,' not potentials from which the forces are derived. As long as one can shift the gauge 
of the potentials without affecting the forces, the description of the system should remain 
the same. But the mass term for the $W^\pm$ gauge bosons, for example, would not remain invariant 
if the gauge of $W^\pm_\mu$ were shifted, unless $m_w=0$. Similarly, a mass term for fermions 
must necessarily mix left-handed and right-handed fermions, but these have different gauge 
quantum numbers, so a shift in gauge would not allow such a term in the Lagrangian density 
to remain invariant. The latter requirement is commonly referred to as chiral symmetry, 
meaning that the Dirac action ought to remain invariant under a chiral rotation.

The addition of a spin-0 scalar field to the standard model introduces an additional interaction 
for the fermions and gauge bosons, regulated by a unique coupling constant $g_j$ for each particle
species ``j'', chosen to produce consistency with the observed masses. The term associated with 
each particle-Higgs interaction appearing in the Lagrangian density is represented as a product
$g_j\xi_j\Phi$, written in terms of $g_j$, the particle field $\xi_j$, and the Higgs field. 
But still nothing interesting would happen with this in terms of generating inertia if all the 
fields retained a zero expectation value in vacuum. This interaction term would then merely vary 
stochastically as the fields fluctuated about zero, and could in no way be linked to the highly 
stable masses we measure for the standard-model particles. To overcome this deficiency, the 
Higgs field is instead assigned a potential, $V(\Phi^\dag\Phi)$, tuned to prevent its 
lowest-energy state from having $\Phi=0$.  This is done by postulating that
\begin{equation}
V(\Phi^\dag\Phi)=-\mu^2\Phi^\dag\Phi+{1\over 2}\lambda(\Phi^\dag\Phi)^2\;,\label{eq:VHiggs}
\end{equation}
with $\mu^2>0$. Does the Higgs field have some as yet unknown `internal' property or
`structure' that produces such a potential? No one knows, but it is not difficult to see 
that, instead of being minimized at $\Phi=0$, $V$ attains its lowest value for the modulus
\begin{equation}
\Phi^\dag\Phi=v^2\equiv {\mu^2\over\lambda}\;.\label{eq:vev}
\end{equation}
The quantity $v$ is known at the Higgs {\it vacuum expectation value}. In other words, if we 
insist on vacuum corresponding to the lowest energy state for such a potential, $\Phi$ cannot 
be zero; it must have a vacuum expectation value consistent with Equation~(\ref{eq:vev}). 
 
This changes the nature of the interaction term completely, because now we may write
$g_j\xi_j(v+\phi_1)=g_j\xi_j v+g_j\xi_j\phi_1$, in terms of the real part of $\Phi$, given
as $v+\phi_1$. Here, $\phi_1$ represents a fluctuation of the Higgs field away from its 
otherwise constant vacuum expectation value $v$. This achieves the principal result
because $g_j\xi_j v$ is a mass term for $\xi_j$, dependent only on $g_j$, $\mu$ and 
$\lambda$. We interpret this result to mean that a fermion or gauge boson (with $g_j\not=0$) 
plowing through the pervasive Higgs field attracts Higgs bosons to itself, and its inertia 
increases in proportion to the mass carried by the latter \cite{Miller:2008}.

But therein lies the crucial caveat. This mechanism is quite different from the QCD interaction
we described earlier. Whereas the latter results from conservation of momentum and the
back-reaction of gluons radiated by accelerated quarks, the Higgs interaction creates inertia
for the standard-model particles by attracting them to massive Higgs bosons. To make this
work, a potential of the form in Equation~(\ref{eq:VHiggs}) is essential, but we don't know
where it comes from. With it, a non-zero Higgs field pervades all of space, very much like 
the aether proposed to mediate the propagation of electromagnetic waves back in the nineteenth 
century. More seriously, though, this ansatz for the Higgs potential includes a quantity $\mu$ 
with dimensions of mass. Indeed, the mass of the Higgs boson in this model is 
$m_{\rm H}^2=2\lambda v^2=2\mu^2$, and it appears as a {\it free parameter}. {\it There is no 
elucidation or explanation for where it comes from.} Yet clearly all of the standard-model 
masses derived with this mechanism are critically dependent on it. 

To summarize, the Higgs mechanism endows standard-model particles with inertia, yet allows
them to still satisfy all of the essential invariances arising from gauge and chiral 
symmetry. But to do so, the Higgs boson must itself already have inertial mass, and we
have no idea where that comes from. And we should not forget that none of these features 
provide us with any elucidation of the complicated structure of quark and fermion masses and
mixings. Why should the particles all have different couplings $g_j$ to the Higgs field?
And where do these values come from? It is fair to say that we have come a long way exploring
the origin of inertia since the nineteenth century, but no one would claim that we fully
understand it yet. And then there's the question of why inertia (or, more likely $\mg$) ought
to be associated with an energy $E=\mg c^2$ ($=mc^2$), which we shall address next.

\section{The Gravitational Horizon in Cosmology}
If we believe the argument that rest-mass energy is more likely to be associated with $\mg$
than some kind of emergent inertia, the next important factor to consider is the source 
of gravity that couples with the particle to produce this energy. Is it other nearby
particles, the laboratory, galaxy or something even bigger? Certainly, no other force
can be involved in this process because, as we have seen, the equivalence principle
works only for gravity. And quite simply, no other force extends meaningfully to large
enough distances to contribute non-negligibly to rest-mass energy. Thus, since the effects 
of gravity are cumulative, one should reasonably expect that all of the cosmic energy 
density in causal contact with the particle must be coupling gravitationally with it and 
contributing to its `rest-mass' energy. But what fraction of the Universe should we include 
in this `causally connected' region?  Fortunately, recent work in cosmology provides us with 
several indispensable clues to answer this question, notably the role played by the 
so-called apparent (or gravitational) horizon in both the interpretation of observational 
measurements and their theoretical foundation \cite{Melia:2018}. 

To avoid any possible confusion, we should reiterate at this stage that the 
question of energy is entirely independent of how inertia arises. In \S~3 we described early 
attempts at explaining inertia based on the influence of distant matter in the Universe and 
found that Mach's principle has never been successfully incorporated into any working theory 
of gravity. Here, we are again invoking an interaction between local particles and the rest 
of the Universe, though it will now become clear that this interaction must be a 
gravitational one. And this gravitational influence is not at all responsible for creating 
inertia but, as we shall see shortly, it appears to be the origin of rest-mass energy.

Standard cosmology is based on the Friedmann-Lema\^itre-Robertson-Walker (FLRW) metric,
describing a spatially homogeneous and isotropic three-dimensional space, expanding or 
contracting as a function of time:
\begin{eqnarray}
ds^2&=&c^2\,dt^2-a^2(t)\left[{dr^2\over (1-kr^2)}+\right.\nonumber\\
&\null&\left.\qquad\qquad r^2(d\theta^2+\sin^2\theta\,d\phi^2)\right]\;.\label{eq:FLRW}
\end{eqnarray}
This metric is written in terms of the cosmic time, $t$, and and comoving spherical
coordinates $(r,\theta,\phi)$, representing the perspective of a {\it free-falling}
observer, analogous to their free-falling counterparts in the Schwarzschild and Kerr
metrics. The expansion factor, $a(t)$, is independent of position, and the geometric 
constant $k$ is $+1$ for a closed universe, $0$ for a flat universe, and $-1$ for an 
open universe. The latest observations \cite{Planck:2018} are telling us that the 
Universe is flat (with $k=0$), so we shall assume this condition throughout this paper.

It is also helpful to introduce the proper radius, $R(t)\equiv a(t)r$, which is 
often used to express changing (or `physical') distances as the Universe expands. 
Sometimes, $R$ is called the areal radius---the radius of two-spheres of 
symmetry---defined in a coordinate-independent way as $R\equiv \sqrt{A/4\pi}$, where 
$A$ is the area of the two-sphere in the given geometry \cite{Nielsen:2006,Abreu:2010}.

In a cosmology based on the FLRW metric, the term `horizon' may refer to (i) the 
`particle horizon,' characterizing the distance particles have traveled towards an 
observer since the big bang, (ii) the `event horizon,' a membrane that separates 
causally connected spacetime events from those that are not, or (iii) several other 
constructs, each with its own customized application \cite{Rindler:1956}. These all 
have their purpose, but as the measurements continue to improve, it is becoming quite 
clear that one particular definition is emerging as the most relevant for interpreting 
the observations---the (imaginary) surface separating all null geodesics receding from 
the observer from those that are approaching. This is how we formally define the apparent 
horizon, $R_{\rm h}$, in general relativity. It turns out, however, that for an isotropic 
Universe (as described by Eq.~\ref{eq:FLRW}), the apparent horizon coincides with the
better known gravitational horizon \cite{Melia:2018,Melia:2007} first identified in the 
Schwarzschild metric, 
\begin{equation}
R_{\rm h}={2GM_{\rm MS}\over c^2}\;,\label{eq:Rh}
\end{equation}
in terms of the Misner-Sharp mass \cite{Misner:1964},
\begin{equation}
M_{\rm MS}\equiv {4\pi\over 3}R_{\rm h}^3\,{\rho\over c^2}\;,\label{eq:MMS}
\end{equation}
where $\rho$ is the total energy density in the cosmic fluid. 

We must be very clear about what this definition actually means, so let us take a moment 
to carefully dissect it. It follows the standard practice in general relativity of considering
the source of gravity (or, more accurately, the spacetime `curvature') to be the energy 
(in this case $\rho$). But this expression also redefines it in terms of a `gravitational
mass density' ($\rho/c^2$) by tacitly assuming the $E=mc^2$ relation. All the equations that 
follow then have this {\it ab initio} assumption built into them. One can see, however, that
this conversion is merely one of convenience, for $R_{\rm h}$ can be re-written independently 
of $\rho$. Introducing the Friedmann equation,
\begin{equation}
H^2={8\pi G\over 3c^2}\rho\;,\label{eq:Friedmann}
\end{equation}
obtained by putting $k=0$, absorbing the cosmological constant $\Lambda$ into $\rho$ (if 
necessary), and inserting the FLRW metric coefficients into Einstein's equations \cite{Melia:2020},
one can easily combine it with Equations~(\ref{eq:Rh}) and (\ref{eq:MMS}) to show that 
$R_{\rm h}=c/H$, the more familiar expression for the Hubble radius, written in terms of 
the Hubble parameter $H\equiv \dot{a}/a$. Yes, quite interestingly, the empirically derived 
Hubble radius in a cosmic setting turns out to be the apparent, or gravitational, radius. 

The physical nature of $M_{\rm MS}$ first emerged from the pioneering work of Misner and 
Sharp \cite{Misner:1964} on spherical collapse problems in general relativity. It is sometimes 
also referred to as the Misner-Sharp-Hernandez mass, to include the subsequent contribution 
by Hernandez and Misner \cite{Hernandez:1966}. In the cosmic framework, however, $R_{\rm h}$---and
therefore $M_{\rm MS}$---is not static. Unlike the situation with Schwarzschild, in which 
$R_{\rm h}$ is in fact the event horizon, $R_{\rm h}$ in cosmology continues to grow as the 
Universe expands, and may eventually turn into a cosmic event horizon, depending on the 
equation-of-state in the cosmic fluid, i.e., it depends on whether or not $H(t)$ eventually 
approaches a constant. In the next section, we shall demonstrate that a particle's rest-mass
energy is none other than its {\it gravitational binding energy} to the Misner-Sharp `mass' 
$M_{\rm MS}$. Though $M_{\rm MS}$ grows as the Universe expands, it is the ratio 
$M_{\rm MS}/R_{\rm h}$ (see Eq.~\ref{eq:Rh}) that sets the conversion factor from $\mg$ 
to $\mg c^2$ ($=mc^2$), and this ratio remains constant as the Universe expands.

For the reader with a deeper understanding of general relativity, it may also be helpful to 
mention that the Misner-Sharp-Hernandez mass may not be the only definition one may use to 
specify a `global' mass, though there are several good reasons for choosing it in the context 
of FLRW. First and foremost, it is not at all arbitrary, in the sense that only this definition 
is consistent with the $g_{rr}$ metric coefficient. As a result, $M_{\rm MS}$ is the only mass 
that provides an apparent horizon allowing us to write the FLRW metric in terms of the proper 
radius, $R=a(t)r$, and the ratio $R/R_{\rm h}$, signaling how far the observer is from the 
gravitational horizon (see Eq.~\ref{eq:FLRW2} below). 

In general relativity, it is generally non-trivial to identify a `physical mass-energy' in 
a non-asymptotically flat geometry \cite{Faraoni:2015}. But when the spacetime is spherically
symmetric, as we have with FLRW, other possible definitions, such as the Hawking-Hayward 
quasilocal mass \cite{Prain:2016}, reduce exactly to the Misner-Sharp-Hernandez construct.
The same happens with another example, known as the Brown-York energy, which is defined as 
a two dimensional surface integral of the extrinsic curvature on the two-boundary of a 
spacelike hypersurface referenced to flat spacetime \cite{Chakraborty:2015}.

It is important to emphasize that our derivation of the radius $R_{\rm h}$ is fully 
self-consistent with the established understanding of apparent horizons in general 
relativity, which are generally defined---even for non-spherical spacetimes---by 
the subdivision of the congruences of outgoing and ingoing null geodesics relative 
to the observer. For the simpler case of a spherically-symmetric spacetime, these 
reduce to the outgoing and ingoing radial null geodesics from a two-sphere of 
symmetry \cite{Ben-Dov:2007,Faraoni:2011,Bengtsson:2011,Faraoni:2015}. Of course, 
the FLRW metric is always spherically symmetric, so the Misner-Sharp-Hernandez 
mass and apparent horizons are simply related via the Birkhoff theorem and its 
corollary. With spherical symmetry, the general definition of an apparent horizon 
thus always reduces exactly to Equation~(\ref{eq:Rh}) \cite{Faraoni:2011,Faraoni:2015}. 
Another way to put this is that Birkhoff's theorem and its corollary allow us to 
define a `gravitational horizon' in cosmology which, however, is simply identified 
as the `apparent horizon' even in non-spherically-symmetric systems.

It is clear, therefore, that the apparent horizon $R_{\rm h}$ directly tells us 
which portion of the Universe is gravitationally coupled to the observer.  Its 
observational and theoretical implications have been discussed extensively in both 
the primary \cite{Melia:2018} and secondary \cite{Melia:2020,Faraoni:2015} literature, 
though there is still some confusion concerning its properties. The time-dependent
gravitational horizon is not necessarily a null surface, but is sometimes confused with one. 
Some \cite{Davis:2004,vanOirschot:2010,Lewis:2013,Kim:2018} have suggested that objects beyond 
$R_{\rm h}(t_0)\equiv c/H_0$ are observable today (at time $t_0$), which is not 
correct \cite{Bikwa:2012,Melia:2012,Melia:2013a}. Almost certainly some of this discourse 
is due to a confusion between coordinate and proper speeds in general relativity. The 
former may exceed the speed of light $c$, but there is an absolute limit to the latter, 
whose value must be calculated using the curvature-dependent metric coefficients. A 
misunderstanding of this distinction can lead to claims of recessional speeds exceeding 
$c$, even within the observer's particle horizon \cite{Stuckey:1992}.

An indication of $R_{\rm h}$'s relevance to our interpretation of the data is provided
by the many cosmological observations \cite{Melia:2018d} now pointing to what could only 
be called a very curious {\it coincidence}: the data are telling us that 
$\dot{R}_{\rm h}(t)=c$ \cite{Melia:2020}. Those familiar with the Schwarzschild horizon 
might at first find this similar to what they would see in free-fall towards a black 
hole as they cross its event horizon, which would also at that moment appear to be
approaching them at speed $c$. But as we have pointed out, $R_{\rm h}$ in the cosmic 
context is not yet an event horizon (and may never turn into one), so it evolves in 
time at a rate dependent on the equation-of-state in the medium. Yet somehow, the
observations are telling us that $R_{\rm h}=ct$ as a function of cosmic time $t$.

From a theoretical perspective, we know that the gravitational horizon in the cosmic 
setting expands linearly with time only if the cosmic fluid satisfies the zero active 
mass condition from general relativity, i.e., if its total energy density, $\rho$, and 
pressure, $p$, satisfy the constraint $\rho+3p=0$. One can easily understand this from 
the second Friedmann equation, more commonly referred to as the Raychaudhuri equation 
\cite{Ray:1955},
\begin{equation}
{\ddot{a}\over a}= -{4\pi G\over 3c^2}\left(\rho+3p\right)\;,\label{eq:Ray}
\end{equation}
from which one finds that $\ddot{a}=0$ as long as $p=-\rho/3$.

A considerable amount of work has been expended over the past decade trying to understand why 
the Universe would evolve in this manner, and there are now clues---both observational and
theoretical---pointing to some possible explanations \cite{Melia:2020,Melia:2019a}. Insofar
as the topic of this paper is concerned, it is not essential for us to dwell on the
details right now, but it turns out that whether or not $R_{\rm h}$ equals $ct$ is of 
utmost importance to the identification of rest-mass energy as a gravitational binding 
energy. As we shall see shortly, this interpretation works only if $R_{\rm h}$ is indeed 
expanding linearly with time.

\section{Gravitational `Binding Energy' and the origin of $E=mc^2$}
The notion that an influence in cosmology ought to be restricted by a gravitational horizon 
is not easy to grasp because spatial flatness in the FLRW metric (Eq.~\ref{eq:FLRW}) suggests 
the Universe is infinite. But we have to remember that the {\it relative} gravitational 
acceleration between two given spacetime points in the cosmic setting is due solely to the 
energy in the {\it intervening} medium. The Birkhoff theorem \cite{Birkhoff:1923} and its 
corollary \cite{Weinberg:1972,Melia:2007} help us to understand why every observer 
or particle---no matter where they are in the presumably infinite cosmos---is 
surrounded by a gravitational horizon a proper distance $R_{\rm h}=c/H$ away. 
Isotropy ensures that the rest of the Universe outside of a `spherical shell' at 
$R_{\rm h}$ has zero influence on the interior. To be clear, this does not mean that the 
Universe possesses just a single spherical region bounded by $R_{\rm h}$. There exists 
such a horizon centered on every observer, and every particle within the cosmic fluid. 
One should therefore expect such a restriction on the size of a causal region to have 
a significant impact on fundamental physics, especially the question concerning the 
origin of rest-mass energy. All of our discussion thus far points to the gravitational 
interaction between $\mg$ and the gravitating energy lying within the particle's horizon 
$R_{\rm h}$ as the likely source of rest-mass energy. In this section, we prove 
this to be true so long as $R_{\rm h}$ is expanding linearly with time---which appears 
to be what the observations are telling us. 

For reasons that will become clearer shortly, it will be helpful for us to complement the FLRW
metric in Equation~(\ref{eq:FLRW}) with its alternative form written in terms of the observer's 
`physical' coordinates, which include the proper radius $R(t)=a(t)r$. The distinction between 
these two descriptions is that fixing the comoving radius $r$ nevertheless still permits
the proper distance $a(t)r$ to change, whereas the observer may choose to keep the physical 
distance fixed by setting $R$ equal to a constant. It is not difficult to show 
that \cite{MeliaAbdelqader:2009,Melia:2018}
\begin{eqnarray}
c^2\,dt^2-a^2\,dr^2&=&\Phi\left[c^2\,dt^2-\Phi^{-1}\,dR^2+\right.\nonumber\\
&\null&\left. 2c\,dt\,\left({R\over R_{\rm h}}\right)\Phi^{-1}\,dR\right]\nonumber \\
&=&\Phi\left[c\,dt+\left({R\over R_{\rm h}}\right)\Phi^{-1}\,dR\right]^2-
\Phi^{-1}dR^2\qquad\label{eq:trans}
\end{eqnarray}
where, for convenience, we have introduced the function
\begin{equation}
\Phi\equiv 1-\left(\frac{R}{R_{\rm h}} \right)^2\;,\label{eq:Phi}
\end{equation}
which signals the dependence of the metric coefficients $g_{tt}$ and $g_{RR}$ on the
proximity of $R$ to the apparent horizon $R_{\rm h}$.

If we now consider the worldlines of observers that have $t$ as their proper time 
from one location to the next---essentially, the comoving observers---then we may
introduce the proper speed $\dot{R}\equiv dR/dt$ in the line element 
and complete the square in Equation~(\ref{eq:trans}). The FLRW metric thus becomes 
\begin{equation}
ds^2=\Phi\left[1 + \left(\frac{R}{R_{\rm h}} \right)\Phi^{-1}
{\dot{R}\over c}  \right]^2c^2dt^2 - \Phi^{-1}{dR^2}-R^2d\Omega^2.\label{eq:FLRW2}
\end{equation}
The expert reader will see a similarity of this equation with that used to derive the 
Oppenheimer-Volkoff equations for the interior of a star \cite{Oppenheimer:1939,Misner:1964}. 
The latter is static, however, whereas both $R(t)$ and $R_{\rm h}(t)$ vary with $t$ in FLRW.

Written in this form, the FLRW metric allows us to see how its coefficients vary as a function
of $R$, but even more importantly, in terms of the ratio $R/R_{\rm h}$. In principle, we
can use it to determine the variation of a particle's characteristics, such as its energy,
with distance from the observer---all the way up to the gravitational horizon \cite{Melia:2019b}. 
Let us define the 4-momentum of a particle
\begin{equation}
p^\mu \equiv (E/c,p^R,p^\theta,p^\phi)\;,\label{eq:4mom}
\end{equation}
written so that the quantity $E$ has units of energy, and $p^j$ ($j=1,2,3$) represent 
the usual spatial components. We do not assume {\it a priori} the relationship between
$E$ and the vector {\bf p}, but insist on $p^\mu$ being a 4-vector. Then, the actual 
physical connection between $E$ and {\bf p} must be given by the invariance of the
contraction $p^\mu p_\mu$ in the spacetime described by Equation~(\ref{eq:FLRW2}).
For the metric coefficients in this line element, one has 
\begin{equation}
\Phi \bigg[1+\left({R\over R_{\rm h}}\right)\Phi^{-1}{\dot{R}\over c}\bigg]^2
\left({E\over c}\right)^2-\Phi^{-1}\left(m\dot{R}\right)^2=\kappa^2,\quad\label{eq:contraction}
\end{equation}
where the invariant contraction $\kappa^2$ is a scalar that we must now uncover. 
Notice that for simplicity and clarity, we have assumed in this expression that the
particle's motion is restricted to the Hubble flow, i.e., that its velocity is purely
radial, with $p^\theta=p^\phi=0$ and 
\begin{equation}
p^R=m\dot{R}\;,\label{eq:pR}
\end{equation}
in terms of the particle's {\it rest mass}, $m$. 

One accustomed to the language of relativity might be tempted to include a time 
dilation factor in Equation~(\ref{eq:pR}), which simply reduces to the Lorentz factor
$\gamma$ in Minkowski space, but that would be incorrect here, because the cosmic time 
$t$, used to infer the speed $\dot{R}$, also happens to be the local {\it proper} time at 
every spacetime point in the medium. Equation~(\ref{eq:contraction}) therefore correctly
yields the dependence of $E$ on the particle's momentum $m\dot{R}$---everywhere in the 
FLRW spacetime, starting at the origin ($R=0$), where the observer is situated, 
all the way to the gravitational horizon at $R=R_{\rm h}$.

To bring out this physical connection between $E$ and {\bf p} more explicitly, let us
re-write Equation~(\ref{eq:contraction}) in the form 
\begin{equation}
E^2={(c\kappa)^2\Phi+(mc)^2{\dot{R}}^2\over
\left[\Phi+\left({R\over R_{\rm h}}\right){\dot{R}\over c}\right]^2}\;.\label{eq:E}
\end{equation}
We interpret this expression to mean that the particle's energy, $E$, is a function of both
its momentum, $m\dot{R}$, and its distance from the observer in the gravitating medium
within $R_{\rm h}$. We first consider what happens at the horizon, where $R=R_{\rm h}$
and $\dot{R}=c$, while $\Phi=0$. Clearly,
\begin{equation}
E(R_{\rm h}) = mc^2\;.\label{eq:Emc2}
\end{equation}
We might find this hardly surprising, except for two critical facts. First of all, the
particle's momentum at $R=R_{\rm h}$ is not zero, yet this expression appears to be giving
us just the rest-mass energy. Second, notice that the value of $E$ in Equation~(\ref{eq:Emc2})
does not come from $\kappa$, which one would naively have assumed ab initio if we had set 
$p^\mu p_\mu=(mc)^2$. Instead, {\it this energy comes from the momentum $p^R$ transitioning to
its relativistic limit,} $p^R\rightarrow mc$, so that $E\rightarrow p^Rc=(mc)c$ in
Equation~(\ref{eq:E}). The contribution from $\kappa$ itself actually gets redshifted away 
completely because $\Phi\rightarrow 0$ when $R\rightarrow R_{\rm h}$. 

The limit $p^R\rightarrow mc$ when $R\rightarrow R_{\rm h}$ follows directly 
from the Hubble law, which says that the expansion velocity is $v=HR$, in terms of the 
Hubble parameter $H\equiv \dot{a}/a$ and proper distance $R$. Thus one may write $v=c(H/c)R$, 
which simply reduces to $v=cR/R_{\rm h}$, leading to the final result given in 
Equation~(\ref{eq:mdotR}) with the definition $p^R\equiv mv$.

This remarkable result tells us that the observer sees the particle's energy approach
what they can only interpret as an `escape energy' upon reaching the gravitational 
horizon, and this quantity is exactly what they would normally consider to be its
rest-mass energy $mc^2$. One must emphasize the phrase `escape energy' in this conclusion,
because this $E$ is entirely due to the momentum $p^R=mc$ the particle needs to overcome
its gravitational confinement within $R_{\rm h}$. There is no contribution at all to $E$
from $\kappa$ at $R=R_{\rm h}$.

At any other radius $R<R_{\rm h}$, the particle's momentum may be written
\begin{equation}
m\dot{R}=mc\left({R\over R_{\rm h}}\right)\;.\label{eq:mdotR}
\end{equation}
Equation~(\ref{eq:E}) may thus be re-written as
\begin{eqnarray}
E(R)^2&=&(mc^2)^2\left[1-\left({R\over R_{\rm h}}\right)^2\right]\left({\kappa\over mc}\right)^2+\nonumber \\ 
&\null&\qquad\qquad (mc^2)^2\left({R\over R_{\rm h}}\right)^2\;.\label{eq:E2}
\end{eqnarray}
For most FLRW cosmologies, $R/R_{\rm h}$ would be a function of time. Thus, $E$ in 
Equation~(\ref{eq:E2}) could not remain constant at any fixed radius $R$, regardless 
of what value $\kappa$ has. Even so, this energy has the very interesting limit $E\rightarrow c\kappa$ 
when $R\rightarrow 0$, but gives no indication of what $\kappa$ should be.  Our argument relating
rest-mass energy to the gravitational binding energy within $R_{\rm h}$ therefore does not
appear to work very well for arbitrary FLRW metrics. 

The situation changes dramatically for a gravitational horizon expanding at lightspeed,
however, which is what the observations seem to be telling us today. In that case, both 
$R$ and $R_{\rm h}$ scale linearly with $t$, and the righthand side of Equation~(\ref{eq:E2}) 
is entirely independent of time. This is also true of the $g_{tt}$ and $g_{RR}$ coefficients 
in Equation~(\ref{eq:FLRW2}), which means that energy is conserved along the worldlines of 
these particular (comoving) observers \cite{Weinberg:1972,Killing:1892}. An easy way to 
understand this is that a Universe with a linearly expanding $R_{\rm h}$ has zero active 
mass (see \S~IV), so that everything within the gravitational horizon experiences zero net 
acceleration. The particle therefore cannot gain or lose energy from the background as 
the Universe expands. For this special case---and only this one---the energy $E$ in 
Equation~(\ref{eq:E2}) must thus be constant, which therefore means that $\kappa=mc$. 
Then we see that
\begin{equation}
E=mc^2\label{eq:emc2}
\end{equation}
everywhere and at all times.

This is a second remarkable result. It tells us that the particle's total energy $E$
remains constant, independent of $R$, even though its momentum $p^R$ transitions from 
zero at the origin to a maximum $mc$ at $R_{\rm h}$. According to the observer at the
origin, the particle thus appears to have a gravitational binding energy $mc^2$ at
their location, which gradually converts into kinetic energy as $R$ increases,
and $E$ eventually becomes completely kinetic, equal to $(mc)c$, when $R\rightarrow 
R_{\rm h}$.  No matter where the particle happens to be, however, its energy never 
deviates from the fixed value $mc^2$. 

A particle with a peculiar velocity, i.e., a non-zero velocity
relative to the Hubble flow, may have non-zero components $p^\theta$ and $p^\phi$
in Equation~(\ref{eq:4mom}), and its radial velocity---which we shall now call 
$\dot{R}_{\rm part}$ to distinguish it from the Hubble velocity $\dot{R}$ in the
denominator---is not necessarily given by Equation~(\ref{eq:mdotR}). It is easy 
to see that, in this more general case, Equation~(\ref{eq:E}) may instead be written
\begin{equation}
E^2={(c\kappa)^2\Phi+(mc)^2{\dot{R}_{\rm part}}^2+(cR)^2\Phi[p_\theta^2+\sin^2\theta
\,p_\phi^2]\over
\left[\Phi+\left({R\over R_{\rm h}}\right){\dot{R}\over c}\right]^2}\;.\label{eq:Egen}
\end{equation}
But again $\Phi\rightarrow 0$ and $\dot{R}_{\rm part}\rightarrow c$ as $R\rightarrow R_{\rm h}$,
no matter the peculiar velocity, so that we recover the same limiting form of the
`escape' energy, $E\rightarrow (mc)c$ at the apparent (or gravitational) horizon.

Near the origin, however, $\Phi\rightarrow 1$ and Equation~(\ref{eq:Egen}) reduces
to 
\begin{equation}
E^2\rightarrow (c\kappa)^2+p^2c^2\;,
\end{equation}
where $p^2\rightarrow (m\dot{R}_{\rm part})^2+R^2[p_\theta^2+\sin^2\theta p_\phi^2]^2$.
We already showed that $\kappa=mc$ leading up to Equation~(\ref{eq:emc2}), which must
be preserved no matter the momentum, since the contraction $p^\mu p_\mu$ is invariant.
And so we recover the well-known Lorentz invariant form of the energy-momentum equation,
\begin{equation}
E^2=(mc^2)^2+(pc)^2\;,
\end{equation}
near the observer. The cosmological principle then ensures that this relation is
the same for every observer throughout the FLRW spacetime. 

\section{Conclusion}
It is important to emphasize the caveat raised above following 
Equation~(\ref{eq:E2}), that the argument we are making in this paper for the 
origin of rest-mass energy works only if $R/R_{\rm h}$ has been independent
of time throughout the Universe's history. That means that $\dot{R}_{\rm h}$
has been constant at the value $c$ from the Big Bang to today. Among the 
strange coincidences in cosmology, the worst of them is the fact that the 
acceleration of the Universe, averaged over a Hubble time, is zero within 
the measurement error. Of course, this does not mean that $\dot{R}_{\rm h}=c$
from one moment to the next, but if this speed varied according to the
prescription of the standard model without the zero active mass condition,
the probability of seeing an average $\langle\dot{R}_{\rm h}\rangle=c$
today is `astronomically' small, effectively zero. In addition, there is
some evidence that the inclusion of zero active mass in $\Lambda$CDM 
may improve its consistency with the data \cite{Melia:2020}. 

Moreover adopting the zero active mass condition appears to eliminate
all horizon problems \cite{Melia:2013b,Melia:2018c}, eliminate the
standard model's initial entropy problem \cite{Melia:2021a}, and 
provide an explanation for how initial quantum fluctuations created 
in the early Universe might have classicalized to produce the 
large-scale structure we see today \cite{Melia:2021b}. If the 
argument we are making here for the origin of rest-mass energy
survives the test of time, perhaps it too may be used to argue in 
favour of zero active mass in the real Universe.

We are justified in calling $mc^2$ the particle's gravitational binding energy because
the observer at the origin infers this to be the energy it needs to reach `escape' 
velocity at $R_{\rm h}$ and free itself from its gravitational coupling to that portion 
of the Universe contained within this horizon. According to the Birkhoff theorem
and its corollary, the rest of the Universe outside of $R_{\rm h}$ does not contribute
to this interaction and is therefore not relevant to the question of rest-mass energy.
Ironically, this interpretation suggests that all particles, those with inertia and
those without, behave equivalently at $R\rightarrow R_{\rm h}$, in the sense that their 
energy there may be written as $E=p^Rc$ in all cases. But whereas the momentum of
massive particles drops to zero from its maximum value, $mc$, at the horizon, that of
massless particles does not change. So while $E=pc$ always represents an energy associated
purely with momentum for the latter, regardless of location, it gradually transitions to a 
`rest' energy associated with $\mg$ ($=m$) for the former when viewed by the observer in 
their vicinity. 

One may wonder how we reached this result without actually having `calculated' the
gravitational binding energy directly. This would be a non-trivial task to
carry out, given that energy in general relativity is not an invariant quantity from
one frame to the next, and would be very difficult to track non-locally. Instead, we
have used the invariance of a contracted 4-vector to do this, which allowed us
to measure the change in the particle's energy (as viewed from the origin) in terms
of its momentum within the Hubble flow. The actual influence of gravity in this
approach is represented by the factor $\Phi(R)$ in the metric. As we have seen,
the redshift effect associated with $\Phi(R)$ accounts for the gravitational
attraction the particle experiences to the rest of the cosmic fluid contained
within $R_{\rm h}$.

A successful interpretation of rest-mass energy as a gravitational binding energy
would lend some support to evidence emerging from cosmological observations 
that the equation-of-state in the cosmic fluid is apparently consistent with the 
zero active mass condition in general relativity. Significant effort is currently
being expended addressing this issue, and the results of this investigation will be 
reported elsewhere. 

{\acknowledgement
I am grateful to the anonymous referee for an excellent, thoughtful 
review of this manuscript, and for suggesting several key improvements to its 
presentation.
\endacknowledgement}

%
%


\begin{thebibliography}{99}
\bibitem{Thomson:1881} J. J. Thomson, Philosophical Magazine {\bf 11} (1881) 229
\bibitem{Stokes:1844} G. G. Stokes, Transactions of the Cambridge Philosophical 
Society {\bf 8} (1844) 105
\bibitem{Englert:1964} F. Englert \& R. Brout, PRL {\bf 13} (1964) 321
\bibitem{Higgs:1964} P. Higgs, PRL {\bf 13} (1964) 508
\bibitem{Heaviside:1889} O. Heaviside, Philosophical Magazine {\bf 27} (1889) 324
\bibitem{Searle:1897} G.F.C. Searle, Philosophical Magazine {\bf 44} (1897) 329
\bibitem{Abraham:1903} M. Abraham, Annalen der Physik {\bf 315} (1903) 105
\bibitem{Lorentz:1892} H. A. Lorentz, Archives N{\'e}erlandaises des Sciences Exactes 
et Naturelles {\bf 25} (1892) 363
\bibitem{Lorentz:1904} H. A. Lorentz, Proceedings of the Royal Netherlands Academy 
of Arts and Sciences {\bf 6} (1904) 809
\bibitem{Melia:2001} F. Melia, {\it Electrodynamics} (University of Chicago Press, Chicago, 2001)
\bibitem{Poynting:1884} J. H. Poynting, Philosophical Transactions of the Royal Society 
of London {\bf 175} (1884) 343
\bibitem{Hasenohrl:1904} F. Hasen\"ohrl, Annalen der Physik {\bf 320} (1904) 344
\bibitem{Hasenohrl:1905} F. Hasen\"ohrl, Annalen der Physik {\bf 321} (1904) 589
\bibitem{Boughn:2011} S. Boughn \& T. Rothman, e-print (arXiv:1108.2250) 2011
\bibitem{Planck:1909} M. Planck, {\it General Dynamics. Principle of Relativity}
(Columbia University Press, New York, 1909)
\bibitem{Einstein:1905} A. Einstein, Annalen der Physik {\bf 18} (1905) 639
\bibitem{Newton:1687} I. Newton, {\it Philosophiae Naturalis Principia Mathematica} (1687)
\bibitem{Yarman:2019} T. Yarman, A. L. Kholmetskii, C. Marchal, O. Yarman \& M. Arik,
Journal of Physics {\bf 1251} (2019) id. 012051
\bibitem{Eotvos:1922} R. v. E\"otv\"os, V. Pek´ar \& E. Fekete, Ann. Phys. (Leipzig) 
{\bf 68} (1922) 11
\bibitem{Dicke:1969} R. H. Dicke, {\it Memoirs of the American Philosophical Society. Jayne
Lecture for 1969} {\bf 78} (American Philosophical Society, Philadelphia, 1970)
\bibitem{Su:1994} Y. Su et al., Phys. Rev. D {\bf 50} 3614 (1994) 
\bibitem{Baessler:1999} S. Baessler et al., Phys. Rev. Lett. {\bf 83} (1999) 3585
\bibitem{Wagner:2012} T. A. Wagner et al., Class. Quantum Grav. {\bf 29} (2012) id. 184002
\bibitem{Tino:2020} G. M. Tino et al., Progress in Particle and Nuclear Physics 
{\bf 112} (2020) id. 103772
\bibitem{Berkeley:1710} G. Berkeley, {\it The Principles of Human Understanding}
(Jeremy Pepyat, Dublin, 1710)
\bibitem{Mach:1872} E. Mach, {\it History and Root of the Principle of the Conservation
of Energy} (English translation published by The Open Court Publishing Co., Chicago 1911)
\bibitem{Einstein:1917} A. Einstein, Sitzungsberichte der K{\"o}niglich Preu{\ss}ischen 
Akademie der Wissenschaften, (1917) pp. 142--152 
\bibitem{Einstein:1946} A. Einstein, {\it The Meaning of Relativity} (Methuen, London, 1946)
\bibitem{Sciama:1953} D. W. Sciama, MNRAS {\bf 113} (1953) 34
\bibitem{BransDicke:1961} C. Brans \& R. H. Dicke, Phys Rev {\bf 124} 925
\bibitem{Pais:2005} A. Pais, {\it Subtle is the Lord: the Science and the Life of 
Albert Einstein} (Oxford University Press, Oxford, 2005) pp. 287--288
\bibitem{Glashow:1961} S. L. Glashow, Nuclear Physics {\bf 22} (1961) 579
\bibitem{Weinberg:1967} S. Weinberg, Physical Review Letters {\bf 19} (1967) 1264
\bibitem{Salam:1968} A. Salam, in {\it Elementary Particle Physics: Relativistic Groups and 
Analyticity. Eighth Nobel Symposium,} N. Svartholm (ed.) (Almquvist and Wiksell, Stockholm, 1968)
\bibitem{Oerter:2006}  R. Oerter {\it The Theory of Almost Everything: The Standard Model, 
the Unsung Triumph of Modern Physics} (Penguin Group, New York, 2006)
\bibitem{Durr:2008} S. D\"urr et al., Science {\bf 322} (2008) 1224
\bibitem{Cheng:1984} T.-P. Cheng \& L.-F. Li, {\it Gauge Theory of Elementary
Particle Physics} (Clarendon Press, Oxford, 1984)
\bibitem{Miller:2008} D. Miller, http://www.scienceinschool.org/print/650 (2008)
\bibitem{Melia:2018} F. Melia, AJP {\bf 86} (2018) 585
\bibitem{Planck:2018} Planck Collaboration et al., A\&A {\bf 641} (2020) A6, 67pp
\bibitem{Nielsen:2006} A. B.~Nielsen \& M.~Visser, CQG {\bf 23} (2006) 4637
\bibitem{Abreu:2010} G.~Abreu \& M.~Visser, Phys Rev D {\bf 82} (2010) 044027, 10pp
\bibitem{Rindler:1956} W.~Rindler, MNRAS {\bf 116} (1956) 662
\bibitem{Misner:1964} C. W.~Misner \& D. H.~Sharp, Phys Rev {\bf 136} (1964) 571
\bibitem{Hernandez:1966} W. C.~Hernandez Jr \& C. W.~Misner, ApJ {\bf 143} (1966) 452
\bibitem{Melia:2020} F.~Melia, {\it The Cosmic Spacetime} (Taylor \& Francis, Oxford, 2020)
\bibitem{Faraoni:2015} V.~Faraoni, {\it Cosmological and Black Hole
Apparent Horizons} (Springer, New York, 2015)
\bibitem{Prain:2016} A.~Prain, V.~Vitagliano, V.~Faraoni \& L. M.~Lapierre-L\'eonard,
CQG {\bf 33} (2016) 145008, 13pp
\bibitem{Chakraborty:2015} S.~Chakraborty \& N.~Dadhich, J High Energy Phys {\bf 2015} (2015) 
id.3, 19pp
\bibitem{Ben-Dov:2007} I.~Ben-Dov, Phys Rev D {\bf 75} (2007) 064007, 15pp
\bibitem{Faraoni:2011} V.~Faraoni, Phys Rev D {\bf 84} (2011) 024003, 15pp
\bibitem{Bengtsson:2011} I.~Bengtsson \& J.M.M.~Senovilla, Phys Rev D {\bf 83} (2011) 044012, 30pp
\bibitem{Davis:2004} T. M.~Davis \& T. H. Lineweaver, PASA {\bf 21} (2004) 97
\bibitem{vanOirschot:2010} P.~van Oirschot, J.~Kwan \& G. F. Lewis, MNRAS {\bf 404} (2010) 1633
\bibitem{Lewis:2013} G. F. Lewis, MNRAS {\bf 432} (2013) 2324
\bibitem{Kim:2018} D. Y. Kim, A. N. Lasenby \& M. P. Hobson, GRG {\bf 50} (2018) id.29, 37pp
\bibitem{Bikwa:2012} O.~Bikwa, F.~Melia \& A.S.H.~Shevchuk, MNRAS {\bf 421} (2012) 3356
\bibitem{Melia:2012} F.~Melia, JCAP {\bf 09} (2012) 029, 10pp
\bibitem{Melia:2013a} F.~Melia, CQG {\bf 30} (2013) 155007, 14pp
\bibitem{Stuckey:1992} W. M. Stuckey, Am J Phys {\bf 60} (1992) 142
\bibitem{Melia:2018d} F. Melia, MNRAS {\bf 481} (2018) 4855
\bibitem{Ray:1955} A. K. Raychaudhuri, Phys. Rev. {\bf 90} (1955) 1123
\bibitem{Melia:2019a} F. Melia, Annals of Phys. {\bf 411} (2019) id. 167997, 5pp
\bibitem{Birkhoff:1923} G.~Birkhoff, {\it Relativity and Modern Physics}
(Harvard University Press, Cambridge, 1923)
\bibitem{Weinberg:1972} S.~Weinberg, {\it Gravitation and Cosmology:
Principles and Applications of the General Theory of Relativity} (Wiley, New York, 1972)
\bibitem{Melia:2007} F.~Melia, MNRAS {\bf 382} (2007) 1917
\bibitem{MeliaAbdelqader:2009} F. Melia \& M. Abdelqader, IJMP-D {\bf 18} (2009) 1889
\bibitem{Oppenheimer:1939} J. R.~Oppenheimer \& G. M.~Volkoff, Phys. Rev. {\bf 55} (1939) 374
\bibitem{Melia:2019b} F.~Melia, IJMP-A {\bf 34} (2019) id. 1950055
\bibitem{Killing:1892} W. Killing, Journal f\"ur die reine und angewandte Mathematik 
{\bf 109} (1892) 121
\bibitem{Melia:2013b} F. Melia, A\&A {\bf 553} (2013) id. A76, 6 pp
\bibitem{Melia:2018c} F. Melia, EPJ-C Letters {\bf 78} (2018) 739
\bibitem{Melia:2021a} F. Melia, EPJ-C {\bf 81} (2021) 234
\bibitem{Melia:2021b} F. Melia, PLB {\bf 818} (2021) id. 136632, 14 pp
\end{thebibliography}
\end{document}